%% file: main.tex
\documentclass[10pt,conference]{IEEEtran}
\usepackage[utf8]{inputenc}
\usepackage{booktabs} % For formal tables
\usepackage[modulo]{lineno}
\usepackage{import}
\usepackage{tabularx}
\usepackage{graphicx}
\usepackage{comment}
\usepackage{array}
\usepackage{subfigure}
\usepackage[numbers]{natbib}
\usepackage{amsmath}
\usepackage{mathtools}
\usepackage{url} 
\usepackage{float}

\input{extra_macros/macro_editing}

\input{extra_macros/macro_tables}

\input{extra_macros/macro_interviews}
\begin{document}

\title{Visualizing test diversity to support test optimisation}

\author{
\IEEEauthorblockN{Francisco G. de Oliveira Neto, Robert Feldt, Linda Erlenhov}
\IEEEauthorblockA{\textit{Dept. of Computer Science and Engineering} \\
\textit{Chalmers $|$ University of Gothenburg}\\
Gothenburg, Sweden \\
\{gomesf,robert.feldt,linda.erlenhov\}@chalmers.se}
\and
%\IEEEauthorblockN{Francisco Gomes de Oliveira Neto}
%\IEEEauthorblockA{\textit{Dept. of Computer Science and Engineering} \\
%\textit{Chalmers $|$ University of Gothenburg}\\
%Gothenburg, Sweden \\
%francisco.gomes@cse.gu.se}
%\and
%\IEEEauthorblockN{Robert Feldt}
%\IEEEauthorblockA{\textit{Dept. of Computer Science and Engineering} \\
%\textit{Chalmers $|$ University of Gothenburg}\\
%Gothenburg, Sweden \\
%robert.feldt@chalmers.se}
%\and
%\IEEEauthorblockN{Linda Erlenhov}
%\IEEEauthorblockA{\textit{Dept. of Computer Science and Engineering} \\
%\textit{Chalmers $|$ University of Gothenburg}\\
%Gothenburg, Sweden \\
%linda.erlenhov@chalmers.se}
%\and
\IEEEauthorblockN{Jose Benardi de Souza Nunes}
\IEEEauthorblockA{\textit{Department of Computing Systems} \\
\textit{Federal University of Campina Grande}\\
Campina Grande, Brazil\\
jose.nunes@computacao.ufcg.edu.br}
}

\maketitle
\thispagestyle{plain}
\pagestyle{plain}

\begin{abstract}
Diversity has been used as an effective criteria to optimise test suites for cost-effective testing. Particularly, diversity-based (alternatively referred to as similarity-based) techniques have the benefit of being generic and applicable across different Systems Under Test (SUT), and have been used to automatically select or prioritise large sets of test cases. However, it is a challenge to feedback diversity information to developers and testers since results are typically many-dimensional. Furthermore, the generality of diversity-based approaches makes it harder to choose when and where to apply them. In this paper we address these challenges by investigating: i) what are the trade-off in using different sources of diversity (e.g., diversity of test requirements or test scripts) to optimise large test suites, and ii) how visualisation of test diversity data can assist testers for test optimisation and improvement. We perform a case study on three industrial projects and present quantitative results on the fault detection capabilities and redundancy levels of different sets of test cases. Our key result is that test similarity maps, based on pair-wise diversity calculations, helped industrial practitioners identify issues with their test repositories and decide on actions to improve. We conclude that the visualisation of diversity information can assist testers in their maintenance and optimisation activities.
\end{abstract}


\input{./introduction.tex}
\input{./diversity.tex}
\input{./methodology.tex}
\input{./results.tex}
\input{./conclusions.tex}

\bibliographystyle{IEEEtran}
\bibliography{references}

\end{document}

%% file: extra_macros/macro_editing.tex
\usepackage[normalem]{ulem} % for \sout
\usepackage{xcolor}

\usepackage{ifthen}
\usepackage{amssymb}
\newboolean{showcomments}
\setboolean{showcomments}{true} % toggle to show or hide comments
\ifthenelse{\boolean{showcomments}}
  {\newcommand{\nb}[2]{
    \fcolorbox{gray}{yellow}{\bfseries\sffamily\scriptsize#1}
    {\sf\small$\blacktriangleright$\textit{#2}$\blacktriangleleft$}
   }
   
  }
  {\newcommand{\nb}[2]{}
   
  }

%% file: extra_macros/macro_tables.tex
\usepackage{booktabs}
\usepackage{array}
\usepackage{colortbl}
\usepackage{multirow}
\usepackage{longtable}

\newcolumntype{v}[1]{>{\raggedright \hspace {0pt}}p{#1}}
\newcolumntype{G}[1]{>{\columncolor{gray90}}#1}

\definecolor{Gray}{gray}{0.8}
\definecolor{gray25}{gray}{0.25}
\definecolor{gray50}{gray}{0.50}
\definecolor{gray75}{gray}{0.75}
\definecolor{gray90}{gray}{0.9}

\newcommand{\grayrow}{\rowcolor{gray90}}

%% file: introduction.tex
\section{Introduction}
\label{sec:introduction}

Several studies report the benefits of diversity-based test case analysis, generation and optimisation~\cite{Feldt2008,Hemmati2015,Noor2015,Feldt2016,Coutinho2016,Henard2016,Miranda2018}. By removing similar test cases, or alternatively generating and selecting dissimilar tests, the resulting test suite is more likely to reveal distinct defects~\cite{Hemmati2015,Cartaxo2011,Feldt2016}. For example, the family of Adaptive-Random Testing (ART) addresses this issues by beginning with a very small test suite, and iteratively generates\slash selects tests more ``distant'' to the current set of tests until certain criteria (typically some variant of coverage, or desired set size or generation time, etc.) are achieved~\citep{Chen2008,Chen2010,Arcuri2011}.

A variety of techniques support diversity-based testing across different domains~\citep{Hemmati2015,Hemmati2013,Cartaxo2011,DeOliveiraNeto2016,Noor2015} and levels of testing (unit~\citep{Feldt2008,Noor2015}, integration~\citep{DeOliveiraNeto2018} and system ~\citep{Tahvili2018,DeOliveiraNeto2016}), and many studies investigate their benefits and drawbacks when used for automated test optimisation such as test case prioritisation or selection. On one hand, these techniques show effective fault detection rate for selective testing~\citep{Cartaxo2011,Hemmati2013,Feldt2016}, while on the other hand, their application can be prohibitive due to costly calculations when used against large sets of test cases or for repeated selection selection~\citep{Arcuri2011,Miranda2018}. Only recently has there been proposals to speed up the required calculations~\citep{Miranda2018}.

Automated diversity-based test optimisation techniques calculate distance values and choosing from (dis)similar tests. Even though there are some proposals of methods that calculate diversity for whole sets of tests at once~\citep{Feldt2016} the vast majority of approaches is based on pair-wise calculations. Not only does this lead to performance challenges, due to the $\mathcal{O}(n^2)$ execution cost, it also makes diversity information hard to visualise and thus to present to developers and testers. Thus it is hard to use diversity information in test analysis and improvement scenarios which involves humans. For example, with a relatively small test suite of a 100 test cases we get 10,000 diversity values with each test case characterized by a 100-dimensional~\footnote{Or really 99 since its distance to itself is trivially 0} vector of distance values. 

When testers and developers cannot be involved in the process or digest the information the results are less likely to get acted upon or have impact~\citep{Feldt2013}. Similar results have been found for debugging: automated tools were disregarded if developers could not trust their results or understand how results had been reached~\citep{Parnin2011}. In addition, diversity-based selection techniques are ultimately limited by the diversity of the original set being selected from and, in certain situations, simple random selection can produce sets with the same, or even superior, defect detection rate~\citep{Arcuri2011,Cartaxo2011,Feldt2017}. Thus, for the full potential of diversity-based approach to be used in test analysis and optimization we need better ways to visualise and work with the many-dimensional, quantitative data it produces. Such techniques could also open up for software engineers and quality assurance staff to compare test suite quality of different systems and build experience over time~\citep{Feldt2013}.

In this paper, we propose to apply information visualization techniques to traditional diversity-based test optimisation results to provide testers with an overview of the diversity of their test artefacts and test sets. Our approach aims to complement the way diversity-based test techniques are used by exposing the diversity information to stakeholders (managers, testers, developers, etc.) in order to support their decisions, such as which test cases to focus the selection on, and which parts of the test suite require maintenance. We evaluate our approach on a case study using three active projects from our industrial partner, a large Swedish company in the retail business with in-house software development and testing activities. By applying the techniques to different types of test information and artefacts we also provide further evidence of the versatility of diversity-based analysis and relative benefits of how they are applied.

Our results show that, in addition to the known benefits of automated test optimisation, visualising the diversity information exposes multiple issues with the investigated test repositories and can be a basis for improvements. Particularly, our interviews with relevant stakeholders at the company reveal that the diversity information helps testers in: i) identifying unexpected levels of redundancy in their test artefacts, ii) guide test maintenance activities by exposing redundant and wasteful test artefacts, iii) allows practitioners to decide when it is beneficial to apply different test optimisation techniques. 
A strength of our study is that we focus on high-level test artefacts, where tests are executed manually and written using natural language. Most previous work focuses on regression test scripts that are automatically executed. 

%A limitation of our study
%On the other hand, our initial results are limited to a selected set of sources of diversity measured with one distance functions (Jaccard Index). 

%% file: diversity.tex
\section{Background and Related Work}
\label{sec:diversity}

\subsection{Diversity-based test optimisation}
\begin{figure}
  \includegraphics[width=0.45\textwidth]{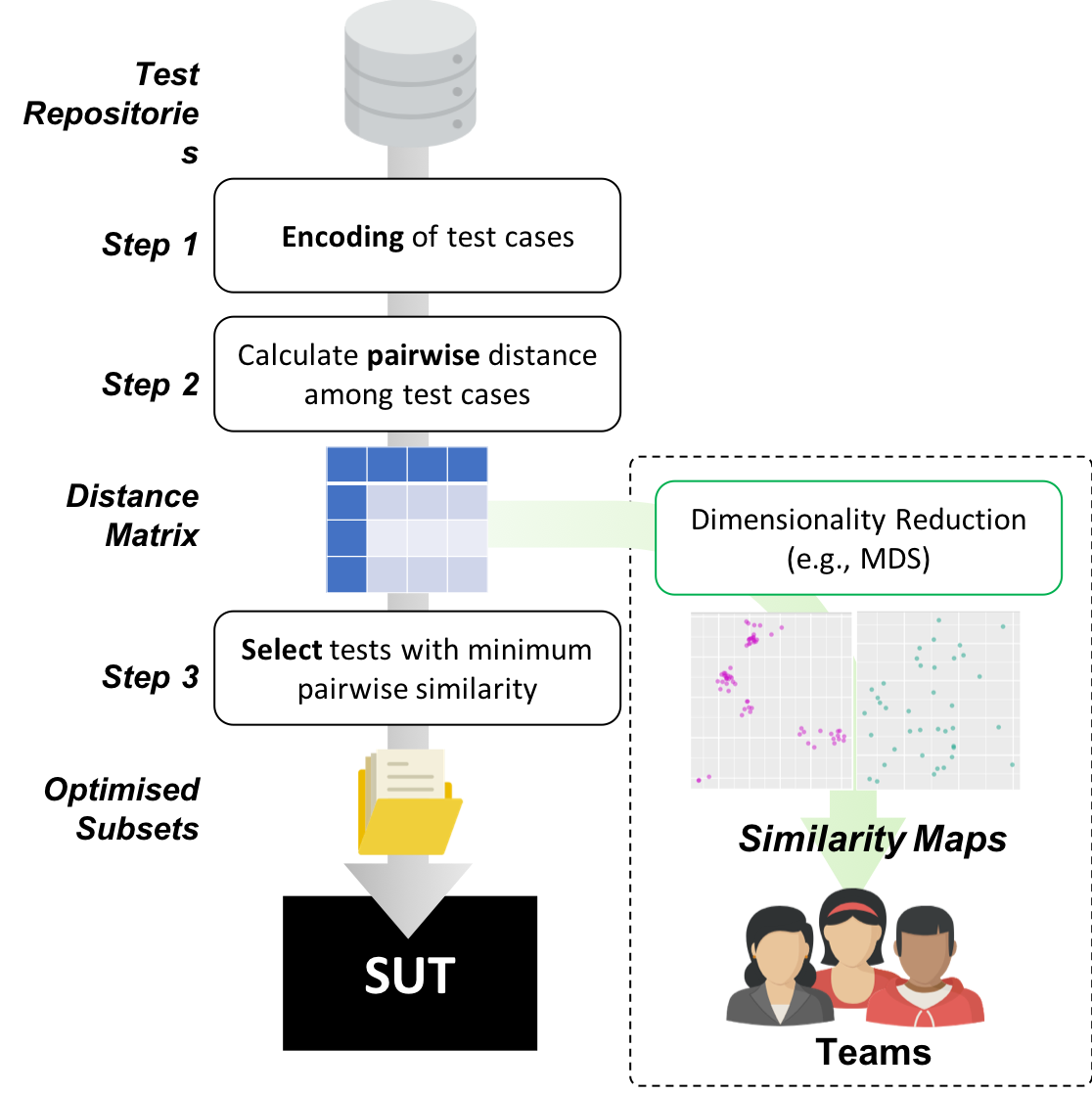}
  \caption{The general steps for diversity-based test optimisation. Our paper proposes the creation of similarity maps and presentation of the diversity information to stakeholders.}
  \label{fig:diversity-optimisation}
\end{figure}

Figure \ref{fig:diversity-optimisation} presents an overview of the steps necessary to perform diversity-based test optimisation. The traditional process includes Steps 1--3~\citep{Hemmati2013}, whereas our contribution proposes an alternative path with a different purpose to the diversity goal. Throughout this section, we focus on test selection, but different types of optimisation (e.g., prioritisation or minimization) can be performed by changing Step 3. Additionally, different techniques are adapted by adding, removing or changing operations between the steps (e.g., hierarchical clustering).

The first step is collecting data from test repositories and encode the diversity information from each test case as a vector (Step 1). One of the main benefits of diversity-based techniques is that they are general and can be applied in different sources of information by choosing different encoding strategies. One can choose to encode static (e.g., test steps, execution traces, code statements) or dynamic (e.g., execution history) information from test artefacts. Consequently, this step requires domain knowledge or an expert's opinion, since the wrong encoding strategy can affect the benefit of the technique. 

For instance, if practitioners choose to encode dynamic history, but do not have enough execution information, the techniques would not be able to capture diversity simply because the existing test artefacts do not have the required information. In our approach, we will use the textual encoding (i.e., strings) of the test artefacts as input for the technique, since test cases and requirements are written manually and in natural language by testers.

In the next step, the pairwise similarity from tests is compared with respect to the encoded information. This step relies on distance functions~\citep{Feldt2008,Ledru2012,Feldt2016} that can quantify the distance between two elements. In other words, the function receives two elements as input (in our case, two test cases) and returns a value indicating their distance. Some functions normalize the distances when used against sets of elements, such that two elements are identical if their distance is zero, or are completely different if their distance is one.

An extensive catalogue of different distance functions can be assembled from the literature~\citep{Nikolik2006,Feldt2008,Hemmati2013,Ledru2012,Feldt2016,DeOliveiraNeto2016}; each of them operate differently depending on what type of element is provided. For instance, the Euclidian distance measures the distance between two points, so the elements are on an interval scale. Here we focus on functions that measure the distance between two strings, considering their lexicographical information. Particularly, we use the Jaccard index (Equation \ref{eq:jaccard}) to calculate the distances between test cases $t_i$ of a test suite $T$, where $i = 1, 2, ..., |T|$.

\begin{small}
\begin{equation}
\label{eq:jaccard}
Jacc(t_i,t_j) = \frac{|t_i \cap t_j|}{|t_i \cup t_j|}\\
\end{equation}
\end{small}

We use Jaccard Index by extracting the test content as a string and converting it into sets of \textit{k-grams} (i.e., sequences of k characters)~\citep{Kondrak2005}. Jaccard Index then operates on those sets to determine the distance between two tests. Even though there are distance functions on strings that have more theoretical support and can be argued to be more general~\citep{Feldt2008,Feldt2016}, the Jaccard index has been widely used in different studies given its simplicity and the positive results it has shown on textual artefacts such as strings~\citep{DeOliveiraNeto2018,Miranda2018,Coutinho2016}.

The distance between all pairs of test cases is then calculated and arranged into a matrix $M_d$, so that $\forall t_i,t_j \in T, a[i,j] = distance(t_i,t_j) | i,j = 1,2,...,|T|.$. Note that: i) the matrix in our distance functions is symmetric, such that $distance(t_i,t_j) = distance(t_j,t_i)$; and ii) we do not use the diagonal, since we are not interested in the distance between a test case and itself. As previously noted, the number of calculations grow significantly for larger sets of test cases, since $|M_d| = |T|^2$.

In Step 3 diversity-based techniques typically iteratively selects from (dis)similar pairs of test cases until a desired criteria is met (e.g., specific subset size, or achieving a coverage threshold). Two common ways of choosing from the matrix is to either: i) start from empty\slash small set and add the most dissimilar pair of test cases~\citep{Miranda2018}; or ii) start from a large set and remove the most similar test cases~\citep{Cartaxo2011,DeOliveiraNeto2016,Feldt2016}. In practice, these two approaches should not significantly change the resulting diversity~\citep{Coutinho2016}, and would only affect which tests are included first in the subset. Alternatively, other approaches use the information to cluster and then select tests~\citep{Tahvili2018,Miranda2018}. The result of Step 3 is a diverse subset of tests $T' \in T$.

During this process the diversity information is typically kept internal to the algorithm\slash tool, while testers only see the resulting subsets\slash clusters. In cases where the test repository would already be diverse enough, then simple random selection could be used for Step 3~\citep{DeOliveiraNeto2018,Cartaxo2011}, or even to avoid the entire process altogether. The main instrument quantifying the diversity of the test suite is the distance matrix, which is not easily perceived by humans since it contains very high-dimensional information.

Therefore, we propose an alternative step that can use methods for dimensionality reduction, e.g. Multidimensional Scaling (MDS)~\citep{Kruskal1964}, to reduce the dimensions of the matrix which allows the visualization of the distances\slash similarities between tests, namely a \textit{test similarity map}. Such similarity maps can then be used during team meetings (e.g., daily stand-up meetings) to discuss the status of the test repository and how to perform certain test activities, such as planning test sessions, or maintaining parts of the test suite.

\begin{figure}
  \centering
  \includegraphics[width=0.45\textwidth]{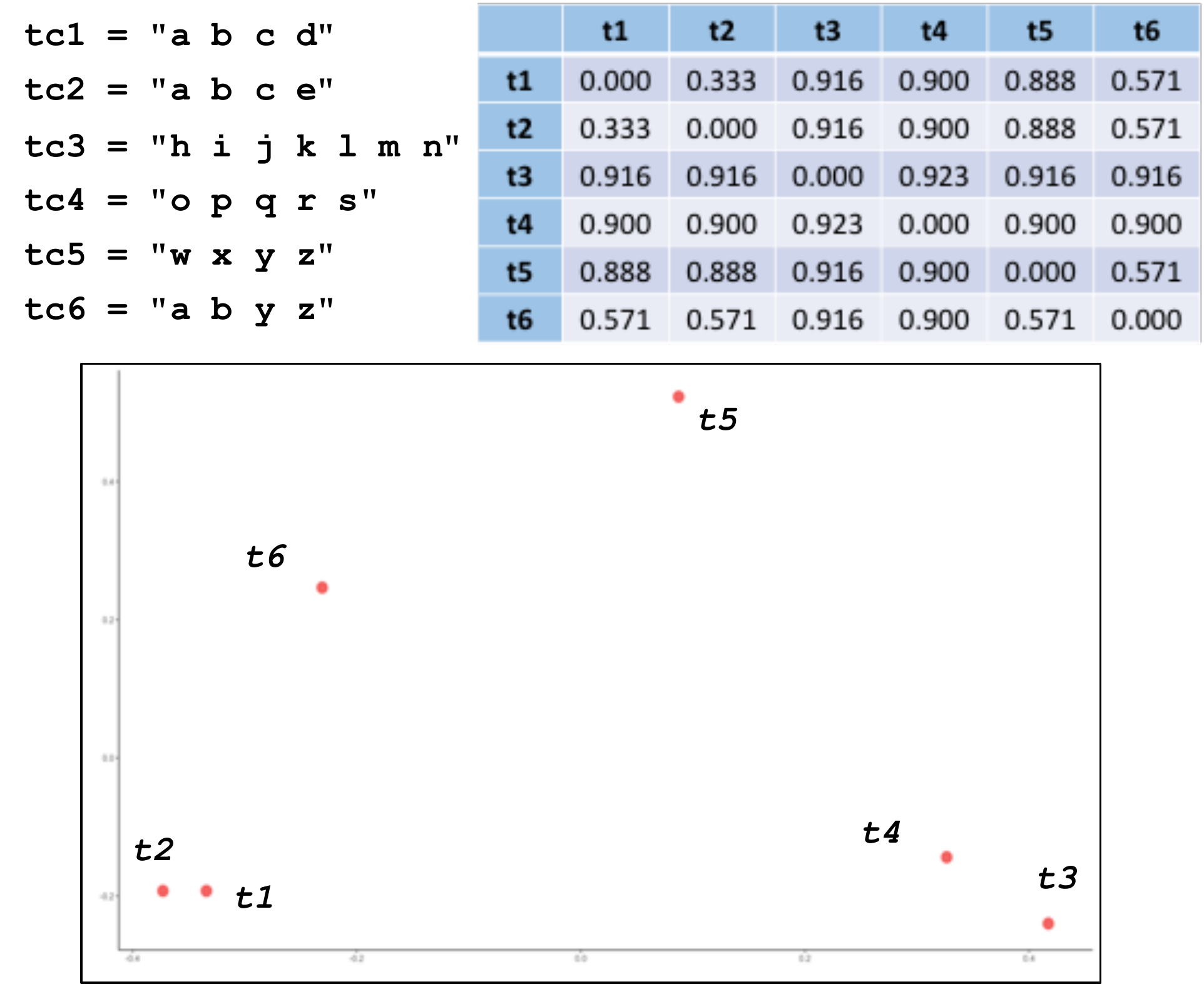}
  \caption{An example of a similarity map obtained from a toy test suite. We use Jaccard Index based on k-grams of length 5 to calculate the distance matrix. Note that the goal with a similarity map is to observe the relative distances between the tests; the scales on the y-axis and x-axis are thus not important.}
  \label{fig:mds-example}
\end{figure}

A dimensionality reduction technique such as MDS receives a distance matrix $M_{|T|\times|T|}$ as input and returns another matrix $M'_{|T|\times|2|}$ represents the coordinates of each element in a space where the between-object distances are preserved as closely as possible\footnote{Actually, the number of output dimensions is often a parameter to dimensionality reduction techniques but are almost always selected to be 2 or 3 for easier presentation to humans}. Figure \ref{fig:mds-example} illustrates a similarity map obtained from a toy test suite. Note that the MDS algorithm is able to preserve the distance between pairs and visually present how tests are diverse with respect to each other. As an example, consider the pair $t1$ and $t2$ that are highly similar, but at the same time are very different to $t4$ and $t3$. An interesting case is $t6$ that is similar to both $t2$ and $t5$, hence being placed between both tests. Certainly, for a test suite with only six test cases, testers can visually compare the pairs of tests and assess their diversity. However, when the number of test cases increase this becomes infeasible. During our interviews with practitioners, they report that for a test suite with more than 10 tests, the similarity maps are a better instrument to convey the diversity of a test suite.

\subsection{Related Work}

Diversity has been a targeted technique to support test optimisation in different domains of testing and with different purposes. The underlying assumption is that faults are located in contiguous regions~\citep{Chen2010,Arcuri2011}, such that similar tests would trigger the same fault. In other words, executing similar tests would not increase fault detection rate, and thus, by selecting\slash generating diverse tests, one can systematically explore the space of possible inputs to search for distinct contiguous area in the space of fault-triggering inputs~\citep{Feldt2008,Chen2008,Cartaxo2011,Arcuri2011}.

The Adaptive-Random Testing (ART) family of techniques, presented the concept of choosing test input based on the current state of diversity of the set of test cases. Therefore, the techniques would need to recurrently calculate the distance between all pairs of tests. Despite the benefits of ART techniques reported in literature, Arcuri and Briand exposed severe drawbacks with the technique when used in realistic scenarios~\citep{Arcuri2011}. Particularly, diversity-based techniques are better used when oracles are available, exhaustive test execution is prohibitive and complex test inputs (e.g., system or integration level) are being used. 

For unit testing that execute quickly, random testing can be a satisfactory candidate as opposed to the costly diversity alternative~\citep{Arcuri2011}. In fact, our findings support this claim even for higher levels of testing, in situations where the test suite has already a satisfactory level of diversity. The challenge then becomes using the appropriate tools to capture that diversity without the prohibitive costs of pairwise distance calculations.

In parallel, several studies investigate the application of diversity-based in different domains, such as model-based testing~\citep{Cartaxo2011,Hemmati2013,DeOliveiraNeto2016}, continuous integration pipelines~\citep{DeOliveiraNeto2018,Hemmati2015}, search-based test generation~\citep{Feldt2017} and at higher levels of testing such as acceptance~\citep{Zhang2018}, system~\citep{Tahvili2018} and integration~\citep{DeOliveiraNeto2018}. Alternatively, studies also focus on investigating the trade-off when using different distance functions in distinct sources of diversity such as use cases~\citep{Coutinho2016}, or modified artefacts~\citep{DeOliveiraNeto2016}. Similar to our case study, most techniques are evaluated in terms of a coverage measure (e.g., code statements, execution traces, requirements) and fault coverage. However, the factors affecting the effectiveness of diversity-based techniques are pervasive to both the different software artefacts (test, code, requirements) and the configuration aspects of the techniques (distance functions, selection strategy, etc.). 

Recent studies adjust those factors in order to overcome the costs in running the techniques in large-scale test repositories. Miranda et al., address this issue by using minhashing and locality-sensitive hashing (LHS) algorithms~\citep{Miranda2018} to, respectively, compress large items into small signatures and reduce the scope of comparison to only a subset of items that are likely to be similar. This is an adjustment of Steps 2 and 3 (Figure \ref{fig:diversity-optimisation}) that allows practitioners to apply diversity-based techniques in previously infeasible situations. But similar to other approaches, the similarity\slash diversity information is discarded after the prioritization is done.

In their initial investigation of using natural-language processing (NLP) techniques to prioritize high-level tests, Tahvili et al.\ \citep{Tahvili2018} observed semantic dependencies between integration-level tests using text analysis. Similar to our approach, authors investigate textual similarities and discuss their solution in terms of a decision support system using the textual dependencies between tests. However, their proposed approach does not explicitly include the tester and similar to existing approaches focuses on optimising the set of test cases.

Our approach is complementary to the pursuit of cost-effective and automated diversity-based test optimisation. In a nutshell, we propose that the information from diversity-based techniques  can be collected, transformed and then presented to testers. Our hypothesis is that the diversity information provides valuable insights to the human in the loop when working with large test repositories.

%% file: methodology.tex
\section{Methodology}
\label{sec:methodology}

We evaluate our approach in a case study with three distinct projects at the IT department of a large-scale retail company in Sweden. The company relies on life-cycle management software to manage the progress of software development processes. However, the projects accumulated years of test artefacts, hindering test activities since deciding which tests to run becomes overwhelming to humans.

Our objectives are two-fold: i) to explore if and how testers select diverse test suites in reality and ii) explore the benefits and drawbacks of exposing the diversity information to stakeholders. Since our industry partner focuses on manual system testing, we investigate these techniques on high-level artefacts as opposed to unit or integration level artefacts.

Previous studies \citep{Arcuri2011,Cartaxo2011,Hemmati2013,DeOliveiraNeto2016,Feldt2016,Miranda2018} investigated cost-effectiveness of diversity-based test optimisation in terms of defect detection rate. Therefore, our contributions aim to complement those findings by further exploiting the diversity information. In the following, we investigate the research questions below:\\

\noindent \textbf{RQ1:} Are testers aware of diversity when selecting tests manually?

\noindent \textbf{RQ2:} How can we use automated diversity-based techniques beyond test optimisation?

\noindent \textbf{RQ3:} Can diversity information assist testers in their test optimization and maintenance activities? If so, how? \\

\subsection{Case company and projects}

Due to NDA restrictions we cannot disclose information about the projects and the partner company. Nonetheless, this section presents an overview of the projects under analysis and descriptive statistics about the test artefacts. We selected 3 projects by interviewing senior practitioners at our industry partner. Our criteria for selection were projects: i) using life-cycle management tools instrumented with APIs for automated data collection, ii) had at least one year of testing activities, iii) would have big, yet varied, sizes of test repositories.

All selected projects follow the same test process defined by the company, where roles, test activities and artefacts should be supervised by test managers. We created tools that mine their test repositories for information on test specifications (i.e., test cases), information on test runs (dates, test result, etc.) and high-level requirements connected to the test cases. Test cases are connected to system's requirements through a many-to-many relationship, i.e. a single requirement can be connected to many tests \textit{and} vice-versa. Table \ref{tab:dataset} presents summarized data about the investigated projects.

\begin{small}
\begin{table}
\centering
\caption{Summary of the projects being investigated in our case study. The project activity column indicate the investigated years where the test repository was being used. Additionally, note that the tests are manually executed, so they are not necessarily executed everyday.}
\label{tab:dataset}
\begin{tabular}{l|cccc}
\toprule
\grayrow  \textbf{Projects} & \textbf{\#Tests} & \textbf{Requirements} & \textbf{Executions} & \textbf{Project activity}\\
\hline
Project 1 & 753 & 74 & 1232 & 2014--2017\\
\grayrow Project 2 & 1247 &1326 & 12058 & 2009--2017\\
Project 3 & 3248 & 781 & 8346 & 2012--2017\\
\bottomrule
\end{tabular}
\end{table}
\end{small}

The test specifications are manually written and executed by a human tester that interacts with the System Under Test (SUT). Therefore, each test specification has several test steps written in natural language containing a sequence of user actions and the corresponding expected result from the system. During each test session, the tester manually chooses a small subset of test cases, since exhaustive execution is prohibitive.

\subsection{Planning and variables}

When investigating diversity-based technique, the first step is to establish what type of diversity should be achieved \citep{Hemmati2013}. Through interviews with practitioners, we decided to focus on three distinct sources of diversity: 

\begin{itemize}
\item \textbf{Requirements:} A short textual description of the system's requirement. Testers often use the requirements to guide their test-related decisions~\citep{DeOliveiraNeto2017}, such that the targeted test session covers the new or modified requirements (e.g., regression testing).
\item \textbf{Test name:} Short name used to identify the test\footnote{This information is not simply an ID. Instead, the name is often a one sentence summary of the test case and/or its purpose, as such it can convey important high-level information and, possibly, be suitable for diversity calculation.}, and specific scenarios under test. For instance, when several test cases should be created for main and alternative scenarios, the testers in the project often create distinct test cases with small variations to their names.
\item \textbf{Test steps:} The collection of steps (user actions and expected outputs) specified in the test specification. The sequence of test steps represent a scenario under test, such that alternative and exception flows end up covering the same (or similar) initial steps.
\end{itemize}

By covering all these three sources of static diversity we cater for distinct levels of granularity perceived by the tester when navigating through the test artefacts. We assume that testers base their decision on more general information (requirements), a more sensible breakdown of that information (test names), or lastly, the detailed sequence of interactions with the SUT (test steps) to understand the underlying elements covered by each test case. Note that for these types of manual test cases there was very little information available on historical test executions and their outcomes. For other scenarios and companies such dynamic information about test cases could be available and then used for creating dynamic test similarity maps. We leave this for future work.

We compare three different techniques: the manually selected test cases (\texttt{Manual}), a prioritized subsets using the Jaccard index\footnote{All distance values are normalized.} (diversity-based prioritization, or simply \texttt{DBP}) and a random subset of tests (\texttt{RDM}). Since our focus is on evaluating the similarity maps we choose to work with only one distance function. Jaccard index is widely used to measure distance between strings and has been a standard baseline treatment throughout different studies in literature~\citep{Miranda2018,DeOliveiraNeto2016,Coutinho2016}.

We use two metrics: i) the level of redundancy and ii) the average percentage of fault detection (APFD). The latter is a widely used metric to assess the fault detection rate of prioritized test suites~\citep{Miranda2018}, since it considers the position of the test case that reveals the $i$-th fault ($tf_i$). Our test artefacts include information about failures, but lack specific fault information, thus we use a variation of the APFD that will consider the position of the test case that reveals the $i$-th failure (i = 1,2,...$|F|$). Consequently, we consider that each test case reveals a unique failure, which would represent to a one-to-one relationship between faults and failures\slash tests.

\begin{small}
\begin{equation}
APFD(T) = 1 - \frac{\sum_{i=1}^{|F|}tf_i}{|T|\times |F|} + \frac{1}{2 \times |T|}
\end{equation}
\end{small}

In turn, we measure redundancy based on the frequency that each word is used within that subset (Equation \ref{eq:redundancy}). Considering that natural language is used to write the test cases, we assume that similar text indicates similar features of the SUT being tested (specially since tests are very close to the actual requirement). In fact, during our initial interviews with practitioners, they stated that testers are encouraged to be concise when writing about different scenarios.

\begin{small}
\begin{equation}
\label{eq:redundancy}
redundancy(T) = 1 - \frac{\#unique\_words(T)}{\#total\_words(T)}
\end{equation}
\end{small}

In addition, we perform a qualitative assessment of the test repositories by using similarity maps as a tool to visually determine whether a selected subset is indeed diverse. The similarity maps are then shown to practitioners and discussed via a focus group interview with five test managers from the company (4 seniors, 1 junior). The interview was composed of open-ended questions and slides where researchers would present results and prompt participates to share their opinion and expertise regarding the results found.

We organized our analysis in two distinct unit of analysis to cater for the quantitative (redundancy and APFD) and qualitative (focus group interview) nature of our methods. Therefore, we present results separately but discuss on them together in order to highlight common findings. A summary of our case study planning is presented in Table \ref{tab:case-study}.

\begin{small}
\begin{table}
\centering
\caption{Our case study planning according to guidelines presented by Runeson et al.~\citep{Runeson2008}.}
\label{tab:case-study}
\begin{tabular}{p{0.28\columnwidth}p{0.62\columnwidth}}
\toprule
Objective & Explore\\
\grayrow The context & Prohibitive high-level manual testing\\
The cases & 3 projects from industry\\
\grayrow Theory & Diversity-based test case prioritization.\\
Research questions &  RQ1, RQ2 and RQ3 \\
\grayrow Methods & Third-degree data collection: \\
\grayrow         & Archival data and metrics\\
Selection strategy & Ongoing projects with large test repositories\\
\grayrow Unit of Analysis 1: & Redundancy and APFD \\
 Unit of Analysis 2: &  Qualitative assessment of diversity\\
                     &  Interview with focus group\\
\bottomrule
\end{tabular}
\end{table}
\end{small}

\subsection{Set up and instrumentation}

\begin{figure}
  \includegraphics[width=0.45\textwidth]{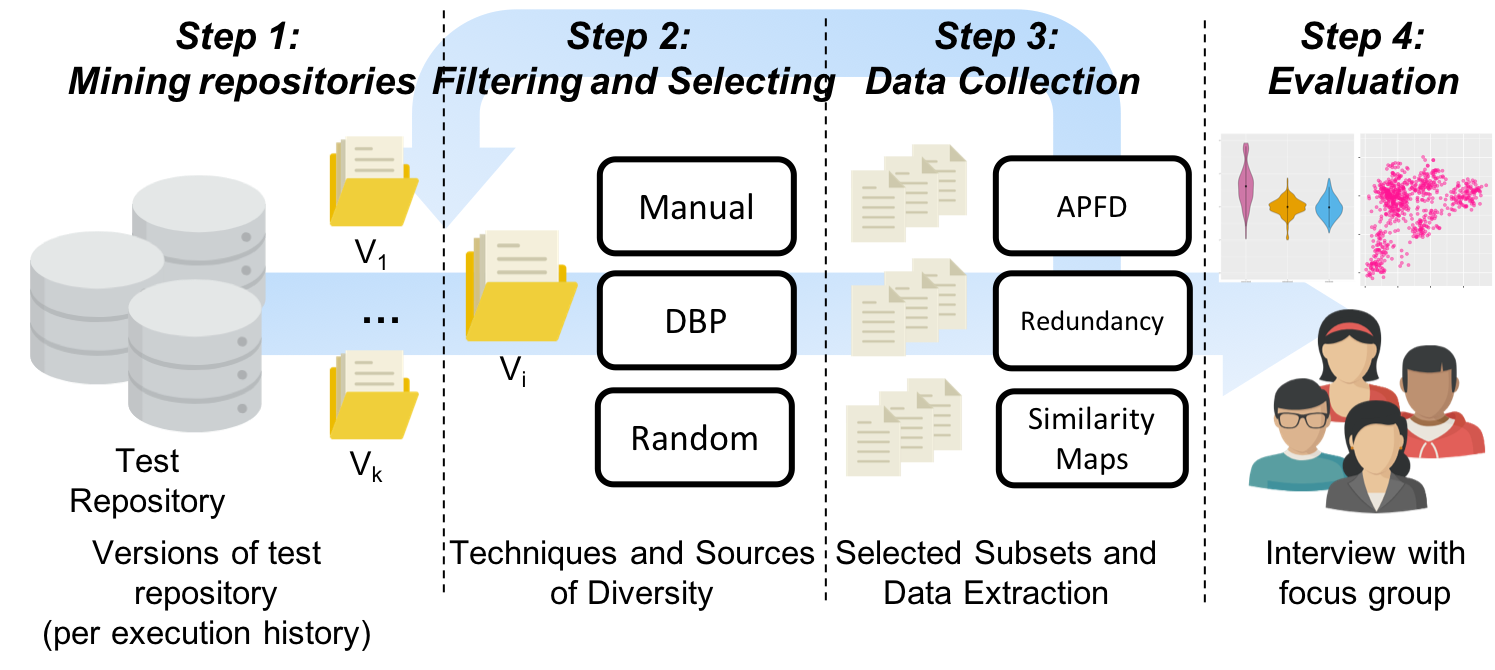}
  \caption{A summary of the steps performed in our case study. In order to compare the selected subsets to the manual selection, we use the corresponding version of the test repository ($V_i$) where the manual selection ($T_{manual}$) was performed. Results were then summarized and presented to practitioners in a focus group interview.}
  \label{fig:summary_study}
\end{figure}

An overview of our method is presented in Figure \ref{fig:summary_study}. We begin by mining the entire test repository and exporting data via APIs provided by the company. We then create different instances of the test repository according to the different dates where a test suite was executed\footnote{In the few cases where several test suites were executed in the same date, we merged all tests into one single test suite}, yielding $V_1, V_2, ..., V_k$, with a corresponding version of the test repository $T_1, T_2, ..., T_k$. Using the information from the test suite selected manually by the tester in that corresponding version $(V_i, T_i)$, we applied two techniques (\texttt{RDM} and \texttt{DBP}) to obtain subsets of test cases with the \textit{same size} as $T_{i,manual}$. In other words, for each registered manual execution we created two subsets of the same size, but instead using automated techniques.

We exclude versions where the manual test suites had fewer than 10 test cases (i.e., $|T_i| > 10$), since we argue that for such small subsets, a human would skip the automated techniques and rely on their own expert opinion to prioritize tests. Moreover, $T_{random}$ was executed 10 times for each $V_i$ resulting in a mean value for both failure and redundancy, whereas the similarity maps are obtained from the 10\textsuperscript{th} execution.

Then, each subset is measured in terms of its own redundancy, APFD and diversity. Note that this step happens three times, one for each source of diversity (Requirements, Name and Steps) addressed in this study, since each will produce a different distance matrix. Lastly, we analyse the data and discuss the results with practitioners in order to answer our research questions. 

We executed the techniques in a MacBook Pro, with $2.2$ GHz Intel Core i7 and 16 GB RAM. The code was implemented in Java and R using open-source libraries to calculate the distance functions\footnote{\url{https://github.com/tdebatty/java-string-similarity}}\footnote{\url{https://cran.r-project.org/web/packages/stringdist/stringdist.pdf}} and perform the MDS \footnote{\url{http://algo.uni-konstanz.de/software/mdsj/}}\footnote{\url{https://stat.ethz.ch/R-manual/R-devel/library/stats/html/cmdscale.html}}. The execution of the DBP took longer for larger test suites depending on the source of diversity used (Table \ref{tab:time}). In fact, Arcuri and Briand raised the scalability issue of diversity-based techniques~\citep{Arcuri2011}, but recent advances presented by Miranda et al. allow testers to overcome it~\citep{Miranda2018}. For our study, scalability was not an issue, but we intend to incorporate Miranda et al.\ strategy to ours in future work.

\begin{small}
\begin{table}
\centering
\caption{Summary of the time required to create the distance matrices.}
\label{tab:time}
\begin{tabular}{l|llll}
\toprule
\grayrow \textbf{Source} & \textbf{Project 1} & \textbf{Project 2} & \textbf{Project 3} \\
\hline
Requirements & 1.64s  & 10s   & 30.1s \\
\grayrow Name         & 6.81s  & 12.6s   &  2.9 minutes \\
Steps        & 2 minutes & 2.3 minutes   & 56 minutes \\
\hline
\grayrow \textbf{Total time:} & \textbf{2.14 minutes} &  \textbf{2.67 minutes}  & \textbf{59.45 minutes} \\
\bottomrule
\end{tabular}
\end{table}
\end{small}

%% file: results.tex
\section{Results and Analysis}
In the next subsections, we first discuss our results in terms of the distinct unit of analysis planned in our case study. We begin with the analysis on redundancy and APFD, followed by a qualitative assessment of our similarity maps with a focus group of practitioners. Then, we relate the results from both units of analysis in order to answer our research questions, followed by discussion on threats to validity.

\subsection{Diversity in test artefacts}
For each treatment, we use violin plots\footnote{Violin plots are similar to box-plots, but more informative, since, in addition to summary statistics, they show the distribution of values.} to present the distributions of redundancy and APFD values for all analysed subsets. In other words, we measure one redundancy\slash APFD value per subset\footnote{As stated in Section \ref{sec:methodology}, we obtain the redundancy for Random by running the technique 10 times (for each $V_i$), and calculate an average.} (i.e., $V_i$) and then show all measured values in the violin plots to analyse the distribution of that redundancy for different selection techniques. On average, testers selected the following number of test cases: Project 1 = 20 ($2.6\%$ of all test cases), Project 2 = 27 ($2.1\%$), Project 3 = 27 ($1\%$).
Several plots showed no significant differences between treatments so, for brevity, we focus on a subset that showed interesting patterns.

\begin{figure}
	\centering
  \includegraphics[width=0.4\textwidth]{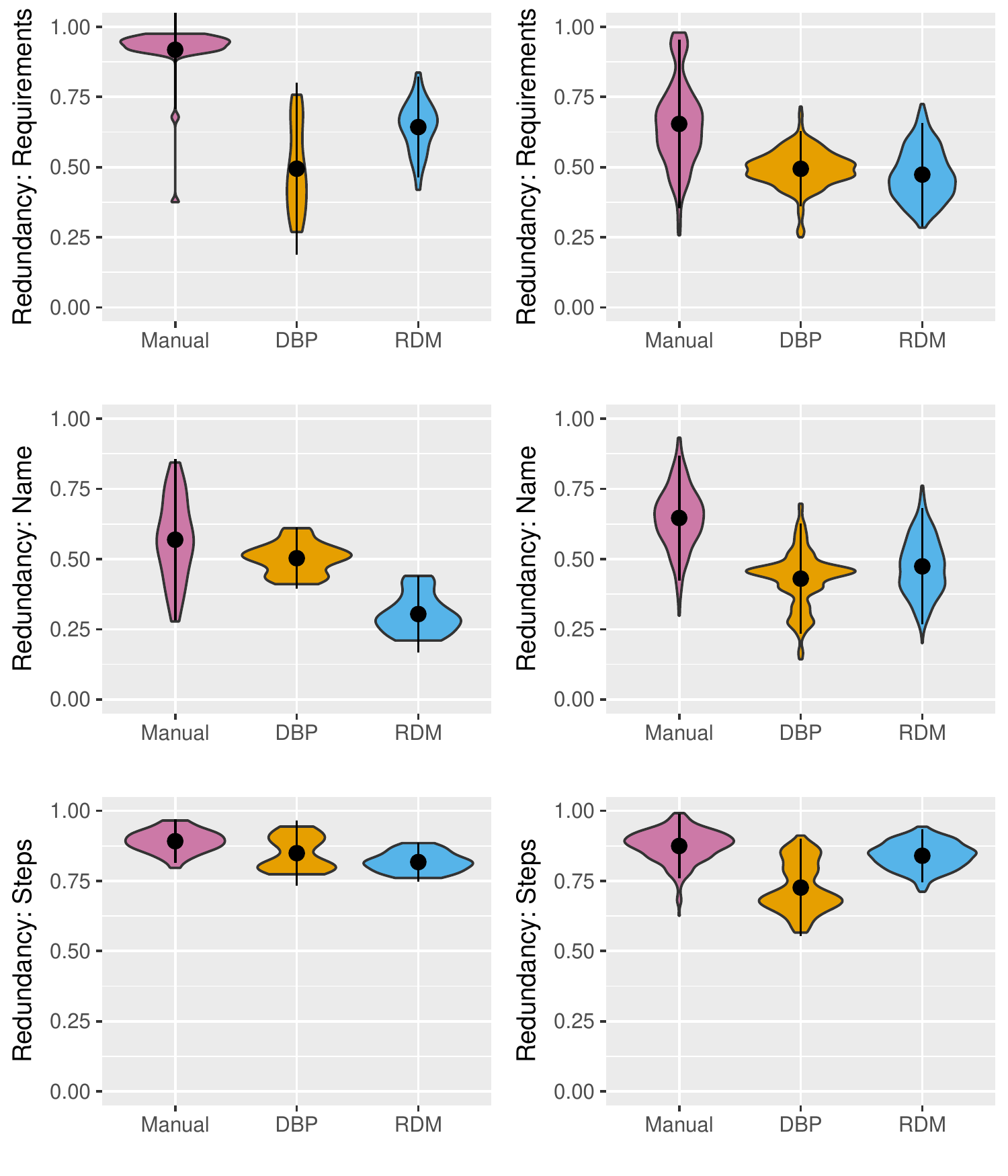}
  \caption{Plots comparing the redundancy results between Project 1 (left) and 2 (right). Note that \texttt{Manual} has higher redundancy in all cases. In addition, \texttt{DBP} has less redundancy in Project 2, while \texttt{Random} is just as good as \texttt{DBM} in Project 1.}
  \label{fig:violin_redundancy}
\end{figure}

Figure \ref{fig:violin_redundancy} presents violin plots for the redundancy values for Projects 1 and 2. Even though there is clear overlap in the distributions for \texttt{DBP} and \texttt{RDM}, \texttt{RDM} gives lower redundancy values for Names and somewhat lower for Steps.

When looking into the full set of diversity values for the test repositories (the numerical values that are the basis for the similarity maps in Figure \ref{fig:sim_maps}), Project 1 is more diverse than Project 2 in requirements, names and steps. Looking at the redundancy plots, it seems that \texttt{DBP} is more beneficial when used in Project 2, and can overcome the project's overall lack of test suite diversity by selecting diverse test subsets which gives lower redundancy values for Name and Steps. However, we see no such \texttt{DBM} advantage for Project 1, which has a more diverse test suite, overall, and \texttt{RDM} can be used instead. Since in both projects small subsets are selected (respectively, 20 and 27 test cases). In other words, it seems that when few tests are sampled and the test repository has a more diverse set of test cases, mostly any randomly selected subset will also be diverse in itself.

\begin{figure}
\centering
  \includegraphics[width=0.45\textwidth]{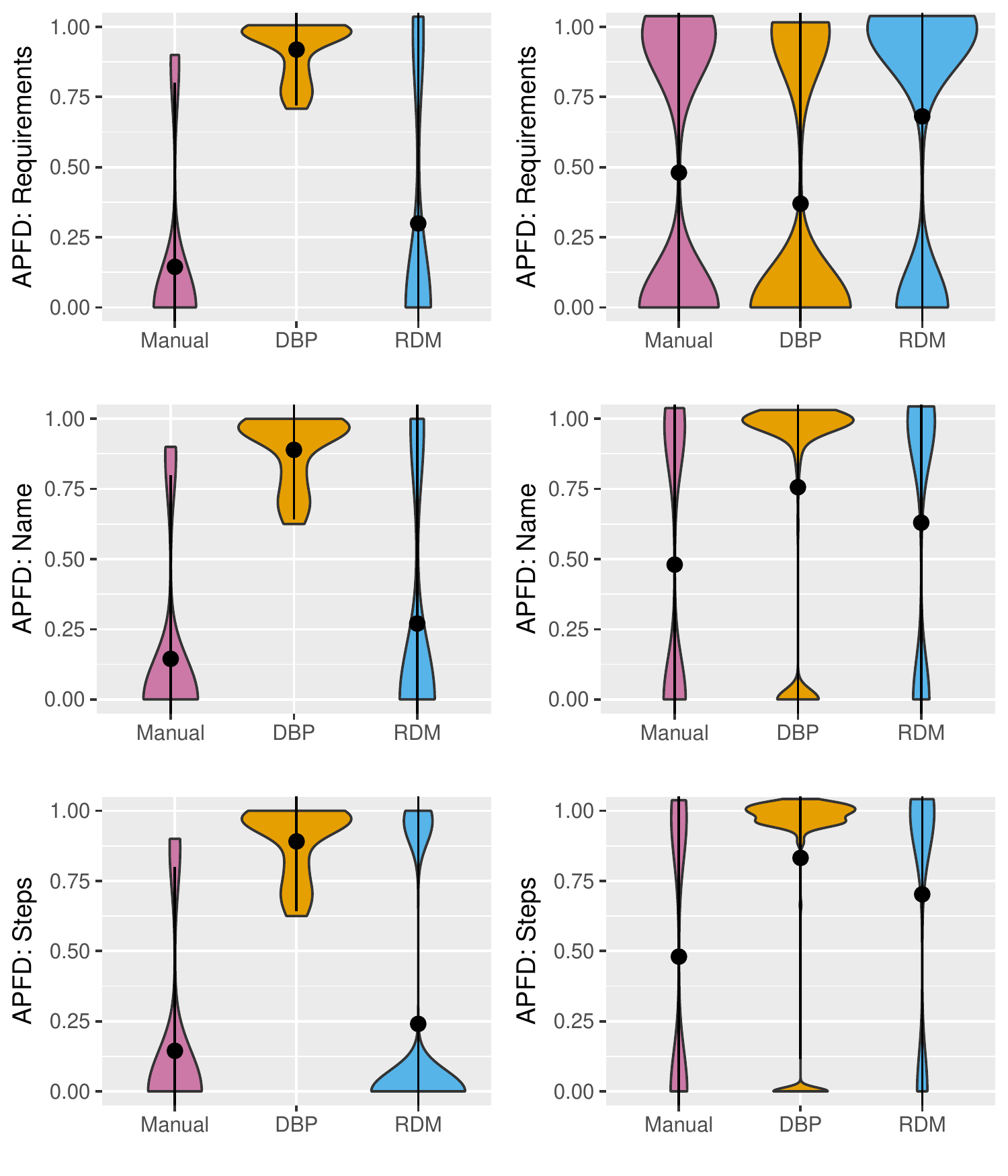}
  \caption{Plots comparing the APFD results between Project 1 (left) and 3 (right).}
  \label{fig:violin_apfd}
\end{figure}

In turn, Figure \ref{fig:violin_apfd} shows the APFD comparison between Project 1 and 3. Note how \texttt{DBP} consistently achieves higher APFD except for the requirements level in Project 3 (top right plot). When looking into the repositories, we identified that Project 3 is also diverse in terms of Name and Steps, however, a large portion of requirements are grouped into the same cluster. Therefore, \texttt{DBM} becomes hindered since it cannot detect diversity information aside from two large clusters of tests. While contrasting this performance with Project 2 (where the lack of diversity actually favours \texttt{DBP}), we observed that Project 2 diversity is still higher than the requirements diversity in Project 3.

We summarize our findings for this part of our results:

\begin{enumerate}
\item Random selection is, in most of our cases, just as adequate as \texttt{DBP} to overcome redundancy in selected test sets. Particularly, we recommend using random selection if the test repository as a whole is already diverse, and only a few test cases are to be selected;
\item Manual selection gives slightly more redundant selection as well as lower APFD scores in all projects and across almost all sources of diversity, and, thus, alternative selection techniques should be considered in industrial practice;
\item The underlying diversity in the test repository, as a whole, can affect efficacy of selection techniques. 
\end{enumerate}

\subsection{Diversity using similarity maps}

\begin{figure}
	\centering
  \includegraphics[width=0.35\textwidth]{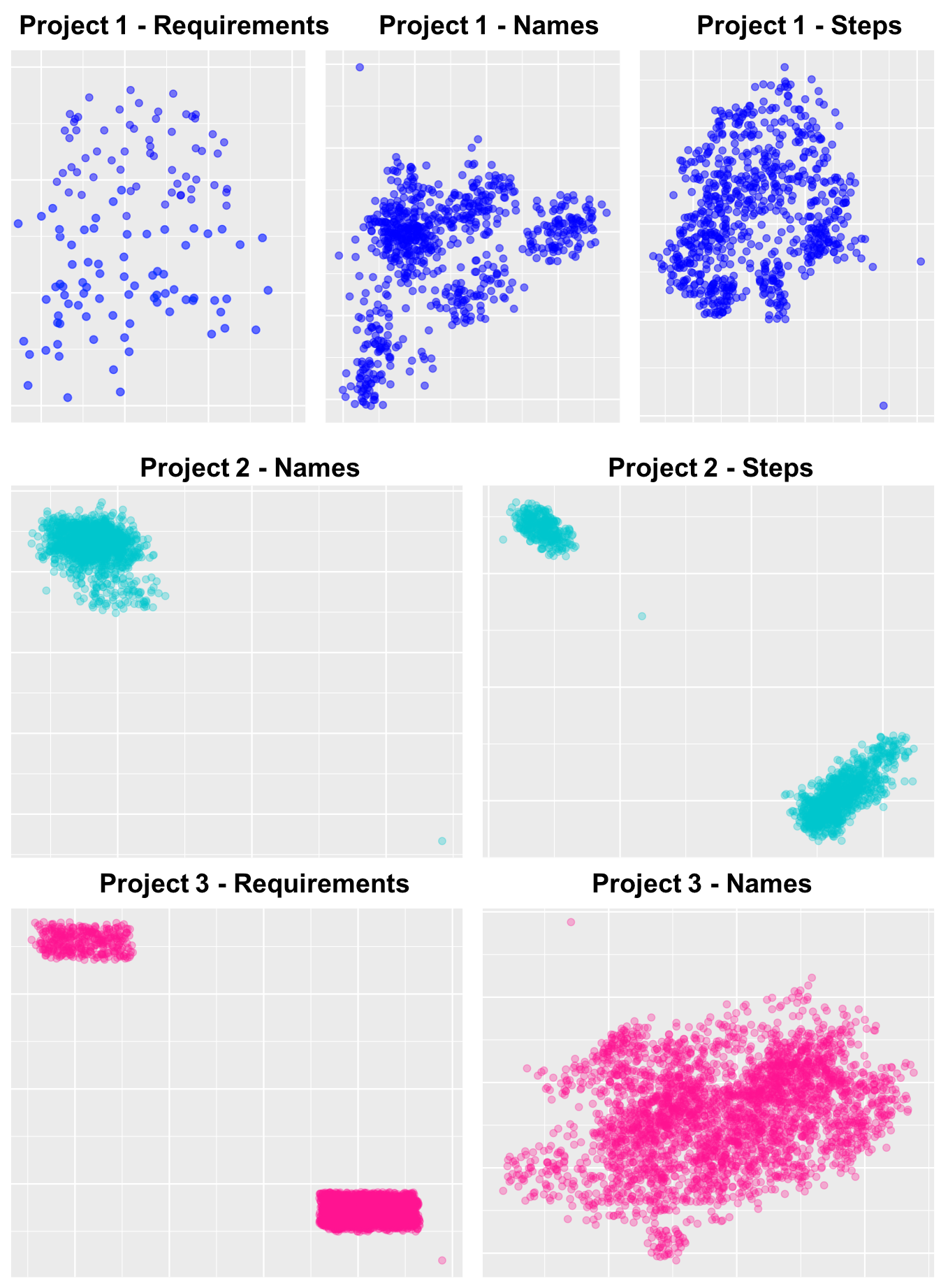}
\caption{Test similarity for full test suites. Only a subset of all 9 maps are shown, for brevity. Darker clusters indicate that a bigger number of tests are grouped.}
    \label{fig:sim_maps}
\end{figure}

\begin{figure*}
    \centering
	\subfigure[]{
    \label{fig:samples_proj1}
    \includegraphics[width=0.31\textwidth]{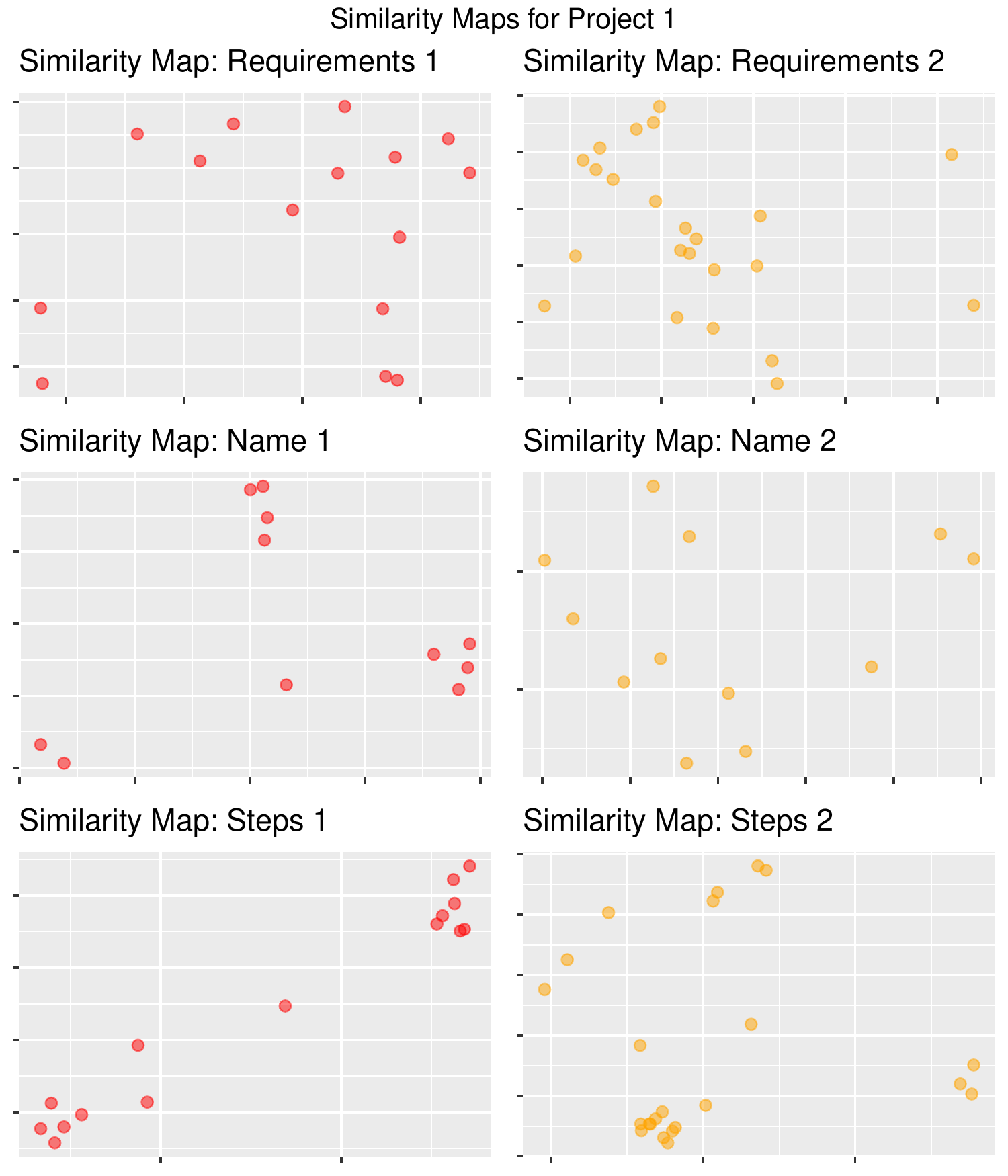}}
	\subfigure[]{
    \label{fig:samples_proj2}
    \includegraphics[width=0.31\textwidth]{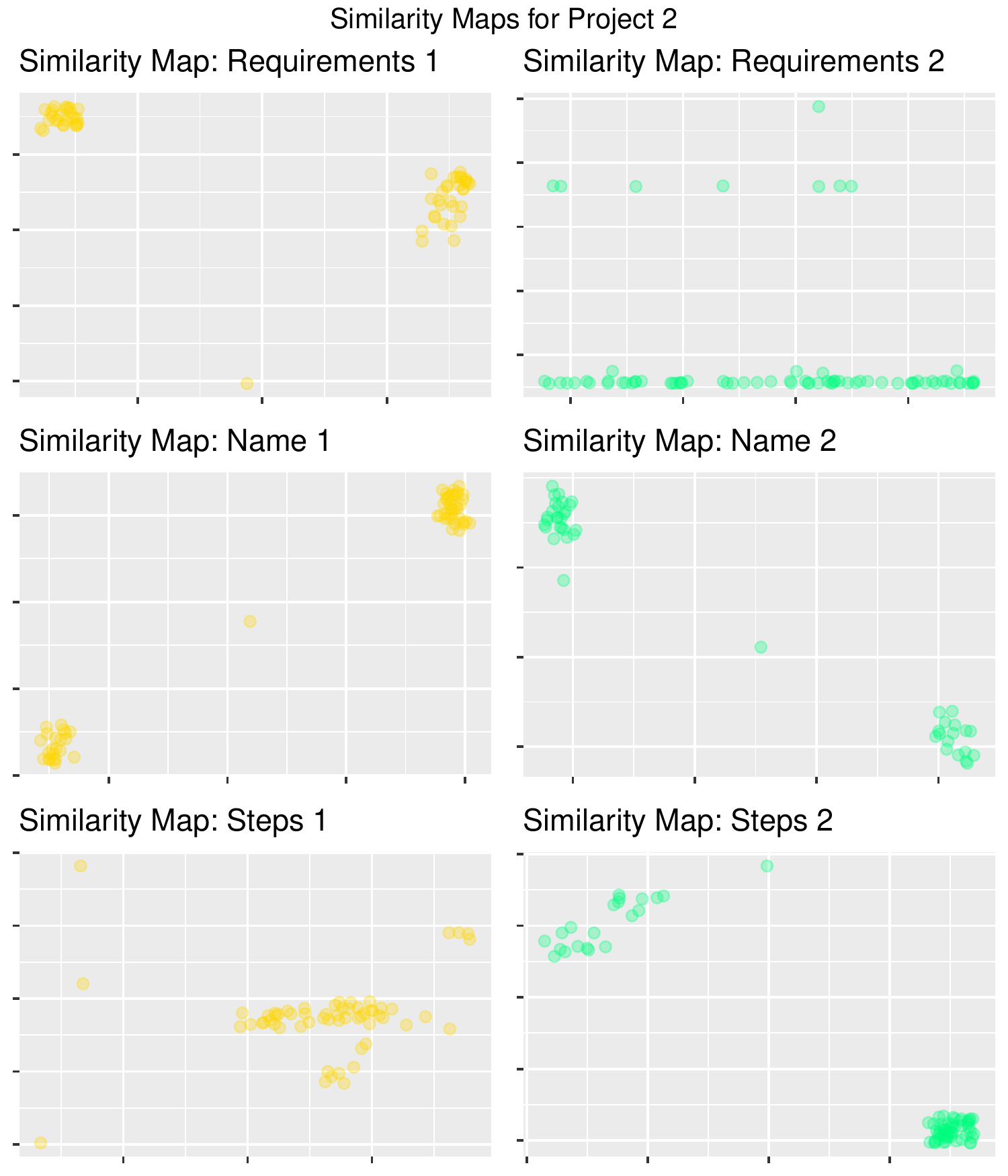}}
	\subfigure[]{
    \label{fig:samples_proj3}
    \includegraphics[width=0.31\textwidth]{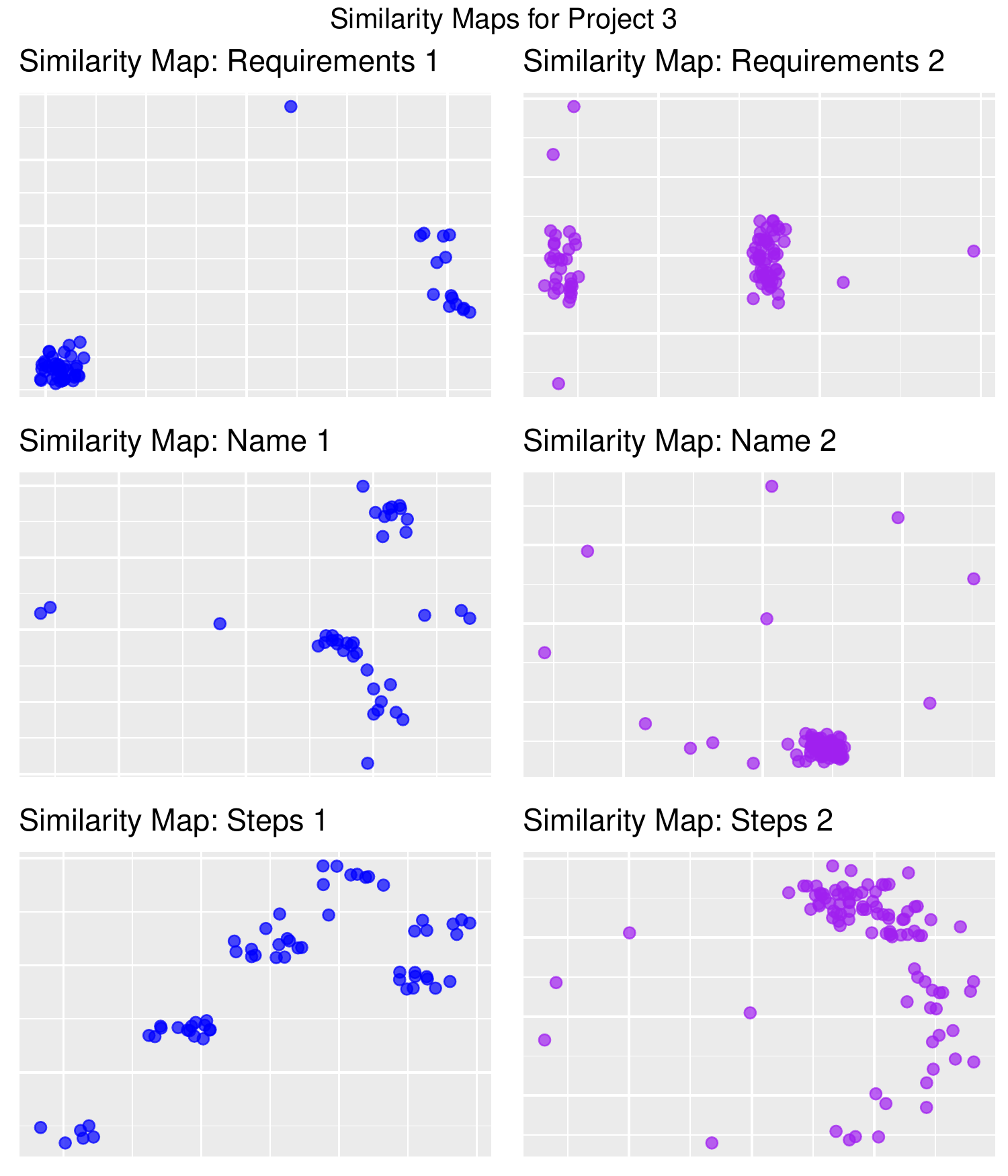}}
    \caption{Similarity maps for all subsets of test cases were generated. We randomly selected 18 (2 per project, and source of diversity) maps presented above.}
    \label{fig:map_samples}
\end{figure*}

As an additional step to the more traditional analysis of diversity above, we created similarity maps using multidimensional scaling using i) the entire test repositories (Figure \ref{fig:sim_maps}), and ii) a sample of manually selected subsets (Figure \ref{fig:map_samples}). The similarity maps in Figure \ref{fig:sim_maps} show how diverse each test repository is. For instance, Project 1 has an overall better spread compared to the other projects. 

The maps were shown to practitioners in order to understand whether the clustering of tests (i.e., similarities among different tests) was intended by testers as well as to provoke discussion. During the focus group discussions, practitioners claimed to be unaware of the large similarities that can be seen for several of the maps. Based on the maps, comparing the maps between projects and when looking into the detailed data for the projects, the senior test managers started to identify problems and consider actions to take. We summarize the insights and suggested action points in Table \ref{tab:insights}.

\begin{table*}
\centering
\caption{Summary of insights and actions based on focus group discussion based on test similarity maps.}
\label{tab:insights}
\begin{tabular}{p{0.005\textwidth}p{0.45\textwidth}p{0.47\textwidth}}
%\begin{tabular}{lll}
\toprule
\grayrow & \textbf{Insights} & \textbf{Suggested action points}\\
\hline
1 & A large percentage of tests could easily be deleted without hindering the test suite in terms of coverage or fault detection rate. & Delete wasteful tests. Similarity maps and diversity values can be used as a starting point for this. \\
\grayrow 2 & Existence of tests that were created but never executed. & Contact the tester assigned to the test and investigate why the test was never executed.\\
3 & Testers create numerous duplicates of tests that do not add value to the process. & Refactor highly similar test scripts to allow reusability of steps instead of redundancy.\\
\grayrow 4 & Specific parts of the repository required immediate maintenance to remove obsolete test scripts. & Either update or remove obsolete test scripts.\\
5 & Some tests did not comply with naming conventions and documenting practices proposed by the test process. & Contact the team and managers responsible in order to encourage compliance with company practices.\\
\bottomrule
\end{tabular}
\end{table*}

In turn, practitioners discussed the diversity shown in specific subsets selected by testers (Figure \ref{fig:map_samples}). One pattern is that test suites that have more diversity allow testers to select diverse subsets. For instance, both illustrated subsets from Project 1 are more diverse than subsets from Projects 2 and 3, in Figure \ref{fig:map_samples}, and overall diversity is higher for Project 1 (Figure \ref{fig:sim_maps}). This pattern was seen across numerous similarity maps created from manually selected subsets.

Another important characteristic is the amount and spread of the clusters themselves. For instance, in Figure \ref{fig:map_samples}, Project 3 has more sparse clusters when compared to Project 2. An alternative course of action is to cater for diversity across different clusters, which is implemented by techniques in literature, e.g., FAST~\citep{Miranda2018}.

In conclusion, the similarity maps revealed meaningful information about the sets of test cases and prompted discussions that lead to numerous insights and improvement possibilities. We observed that the clustering of test cases in the test repository can constrain testers in their choices, specially if they are not aware of the current similarities between parts of the test suite. On the other hand, practitioners stated that navigating through the similarity maps to find specific information on test cases could be cumbersome. We argue that the similarity maps should be used to obtain a holistic view of the diversity status of the test artefacts, whereas other tools (e.g., dashboards in life-cycle management tools) could be combined with the maps to enable in-depth view of specific test cases.

\subsection{Analysis and discussion}

Our findings below are limited to our context of high-level manual testing and textual artefacts written in natural language. Nonetheless, we believe that our results can be applicable to other sources of diversity given the flexibility of diversity-based strategies to easily change between distance functions and to the maps to provide overview and guide discussions and detailed exploration.

\noindent \textbf{RQ1:} Are testers aware of diversity when selecting tests manually?
The manual subsets themselves were more redundant than the ones selected automatically, this indicates they are not currently using diversity as a basis for manual selection. Moreover, our similarity maps show that a diverse test repository should be a prerogative for both manual and automated selective testing. Even though that is not surprising, the long-term value is devise complementary strategy combining automated approaches with regular visual checks for the diversity of the repository.\\

\noindent \textbf{RQ2:} How can we use automated diversity-based techniques beyond test optimisation?
Even though automated optimisation techniques can systematically operate on test diversity, the distance matrix should also be used as an instrument to decide whether to apply diversity-based techniques at all. In fact, automated diversity-based techniques are costly~\citep{Miranda2018,Arcuri2011} and, in our case study, a simple random prioritization can allow satisfactory levels of redundancy and APFD, if the starting diversity is high enough. Other recent research supports this view~\citep{Feldt2017}. Discussion of diversity information can be used to inform test maintenance activities.

\noindent \textbf{RQ3:} Can diversity information assist testers in their activities? If so, how?
During our focus group interview, the similarity maps revealed a series of issues with the test artefacts that practitioners were not aware of. In addition to the overall low levels of diversity for some test suites, practitioners could detect duplicated sets of test cases (or parts of them) that were not being properly maintained (e.g., not updated or deleted from the test repository) by the team members. The maps could provide overview and help direct the discussion and prompt more detailed investigation into particular aspects and data.

In conclusion, the main value of the similarity maps, observed during our study, was to raise awareness and trigger insightful discussions among practitioners during progress meetings. Additionally, one unique aspect of the similarity maps is to allow navigation of the diversity space so that testers can visually find clusters and investigate further issues behind those clusters. That enables teams to identify whether test cases are creating waste in specific sets of test cases. This requires interactive versions of the maps (that we used in the discussions; in this paper PDF we only show static versions, obviously).

\subsection{Threats to validity}

Our construct validity and external validity is limited by the small number of techniques and metrics used to evaluate the test artefacts. Ideally, one should include more distance functions~\citep{Coutinho2016,Feldt2016}, different prioritization heuristics~\citep{Miranda2018}, and consider sources of diversity beyond string representation of static information. We intend to address these threats in future work. Since this is an exploratory case study, we focus on controlling the amount of independent variables in order to attribute benefits and drawbacks to the similarity maps themselves, instead of, e.g., different distance functions or diversity representation (numerical x string).

On that note, the distance functions between strings are only capable of capturing lexicographical information~\citep{Ledru2012}, such that synonyms are not accounted in the diversity. We mitigated the risks above by verifying our redundancy results with the practitioners responsible for the projects under investigation, so that they could identify potential problems with the dataset. In the future, more semantically focused methods, as used, for example, by Tahvili et al~\citep{Tahvili2018}, could help improve this.

In turn, our conclusion validity is limited to descriptive statistics and visual analysis. A detailed distinction between the three techniques (Manual, Random and \texttt{DBP}) could be achieved through hypothesis testing. On the other hand, we argue that the statistical tests would not add to the practical significance of our findings with respect to unit of analysis 1, since. In other words, a statistically significant difference, e.g., between Manual and \texttt{DBP}, would still be subjective to construct validity threats, such as the choice of Jaccard Index. Conversely, our qualitative assessment in unit of analysis 2 with a focus group of practitioners lever the practical significance of our investigation.

In turn, internal validity threats are related to the instrumentation of the study as well as the interview with the focus group. Regarding the former, the techniques were widely tested, including the open-source libraries used in our experiment. We used an interview instrument to interact with the focus group. During the 3 hour session, two of the authors guided the discussions with practitioners during presentation of the results obtained from both unit of analysis. Finally, external validity is limited to our industrial context, which is expected in a case study. Similar to the construct validity threats, we intend to improve external validity in an experimental study by including datasets from other industry partners covering different levels of testing (e.g., integration).

%% file: conclusions.tex
\section{Concluding Remarks}
\label{sec:conclusions}

This paper advances the current practices in diversity-based testing by proposing the use of similarity maps to support stakeholders with test-related decisions and activities. The distance measures between tests are often used for test optimization, hence being discarded after the techniques are executed. We instead leverage diversity values and use information visualization techniques to provide overview graphical views of diversity to testers and other stakeholders. This can trigger insightful discussions and support maintenance and improvement decisions.

We evaluated the use of similarity maps in a case study by contrasting the application of one diversity-based test prioritization technique with manual and random selection on three projects from a company in Sweden. Our analysis reveal that diversity-based prioritization is affected by the overall diversity of the test repository. Moreover, if not checked regularly, the diversity of the entire test repository can erode and, thus, hinder test planning, maintenance, and, ultimately, quality.

Another finding is that simple random selection can be effective when the test repository is already diverse. This was also reported in existing studies, where diversity-based techniques achieve a certain threshold after selecting very small subsets (e.g., less than $10\%$ of the original size of tests)~\citep{Cartaxo2011,DeOliveiraNeto2018}. As a consequence, stakeholders can skip the costs involved in executing~\citep{Arcuri2011} and maintaining~\citep{DeOliveiraNeto2018} automated diversity-based testing in their development cycles.

The exploratory nature of our case study hinders broad generalization of our findings; further experimentation is required. Particularly, we plan to include more diversity-based strategies and more distance functions and to apply test similarity maps to other test information. 